\documentclass{sf2a-conf2023}
\usepackage{graphicx}
\usepackage{hyperref}
\usepackage[]{natbib}  
\usepackage{epstopdf}

\def\BibTeX{{\rm B\kern-.05em{\sc i\kern-.025em b}\kern-.08em
    T\kern-.1667em\lower.7ex\hbox{E}\kern-.125emX}}
\bibpunct{(}{)}{;}{a}{}{,}  

\begin{document}

\TitreGlobal{SF2A 2023}


\title{Citizen Science Time Domain Astronomy with Astro-COLIBRI}
\runningtitle{Astro-COLIBRI}

\author{Fabian Schüssler}\address{IRFU, CEA, Université Paris-Saclay, Gif-sur-Yvette, France}
\author{M. de Bony de Lavergne$^1$}
\author{A. Kaan Alkan$^{1,}$}\address{Laboratoire Interdisciplinaire des Sciences du Numérique, CNRS, Université Paris-Saclay, 91405 Orsay, France}
\author{J. Mourier$^1$}
\author{P. Reichherzer}\address{Department of Physics, University of Oxford, Oxford OX1 3PU, United Kingdom}

\setcounter{page}{237}


\maketitle


\begin{abstract}
Astro-COLIBRI is an innovative tool designed for professional astronomers to facilitate the study of transient astronomical events. Transient events - such as supernovae, gamma-ray bursts and stellar mergers - are fleeting cataclysmic phenomena that can offer profound insights into the most violent processes in the universe. Revealing their secrets requires rapid and precise observations: Astro-COLIBRI alerts its users of new transient discoveries from observatories all over the world in real-time. The platform also provides observers the details they need to make follow-up observations.

Some of the transient phenomena available through Astro-COLIBRI at accessible by amateur astronomers and citizen scientists. Some of the features dedicated to this growing group of users are highlighted here. They include the possibility of receiving only alerts on very bright events, the possibility of defining custom observer locations, as well as the calculation of optimized observation plans for searches for optical counterparts to gravitational wave events.  \end{abstract}

\begin{keywords}
multi-messenger, real-time, follow-up, citizen science
\end{keywords}


\section{Introduction}
\begin{figure}[t!]
\begin{center}
\includegraphics[width= 0.7\textwidth]{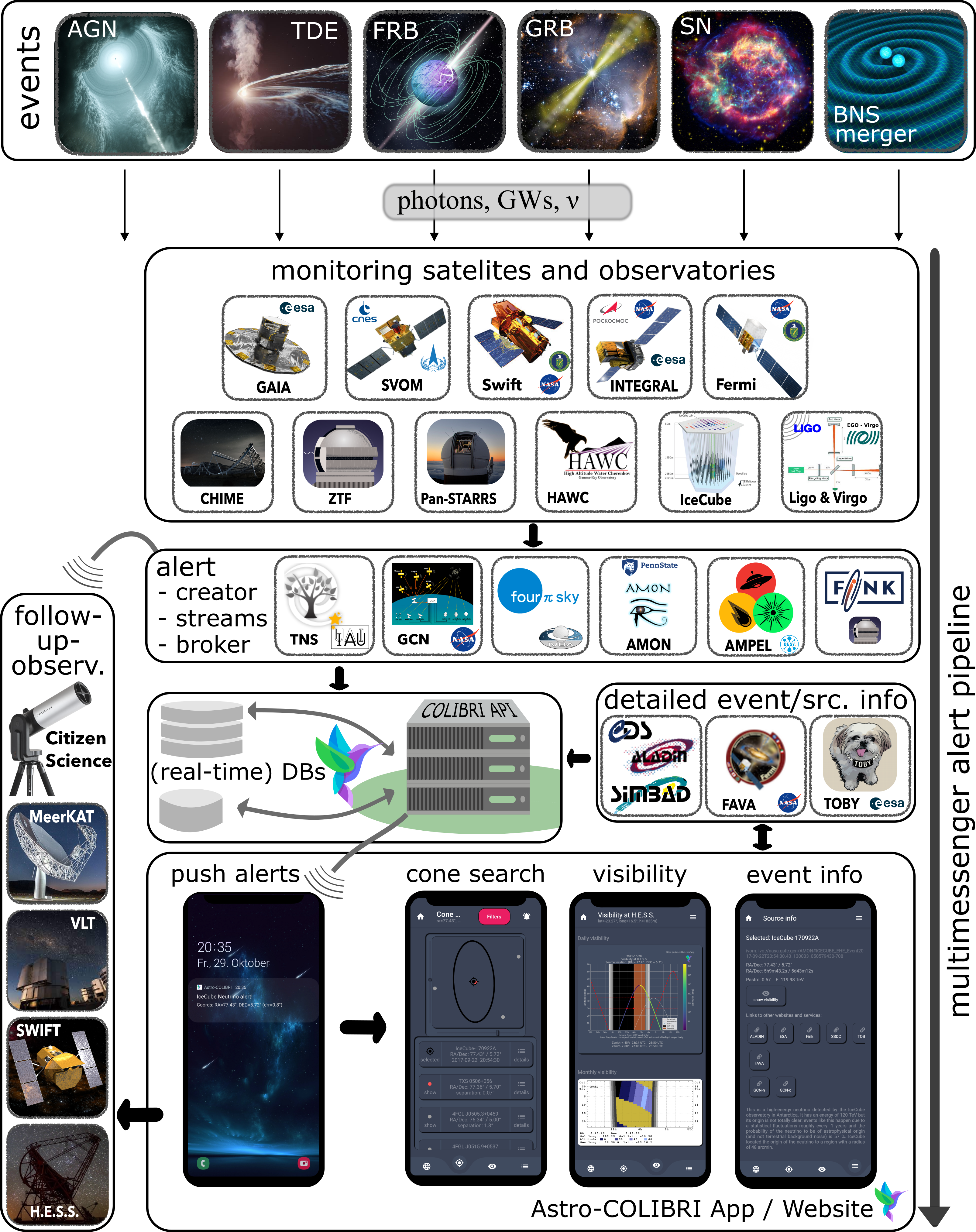}
\caption{Astro-COLIBRI is conveived as top-level platfrom fully integrated in the global multi-wavelength and multi-messenger landscape.}
\label{fig:landscape}
\end{center}
\end{figure}

Astro-COLIBRI~\citep{2021ApJS..256....5R, 2023Galax..11...22R} is a modern system created to help researchers study short-lived astrophysical events. It does this by collecting information about detections of different types of signals in real-time, all in one easy-to-use interface. Astro-COLIBRI brings together the data about these events with tools to plan and faciliate follow-up observations. This helps both professional and amateur astronomers learn more about the violent universe by using different types of observational data. Astro-COLIBRI can be used to study many different phenomena like flaring Active Galactic Nuclei (AGN), Gamma-ray Bursts (GRBs), Fast Radio Bursts (FRBs), Gravitational Waves (GWs), High-energy Neutrinos, Optical Transients (OT) including Supernovae (SN), and more. Aiming to combine information from different observatories and systems, Astro-COLIBRI encourages astronomers to work together and share their data. An overview of the context and place of Astro-COLIBRI fits in the bigger picture of time domain astrophysics is given in Fig.~\ref{fig:landscape}. The system connects to many sources of information that send out alerts about new discoveries and classifications of these short-lived events. It organizes this information and compares it with other time-critical and archival data. Each event is also linked to various external services and expert platforms. Specially designed tools help astronomers get ready to observe the most interesting events. 

\section{Alerts and notifications}
\begin{figure}[t!]
\centering
     \begin{minipage}{0.5\textwidth}
        \centering
        \includegraphics[width=\textwidth]{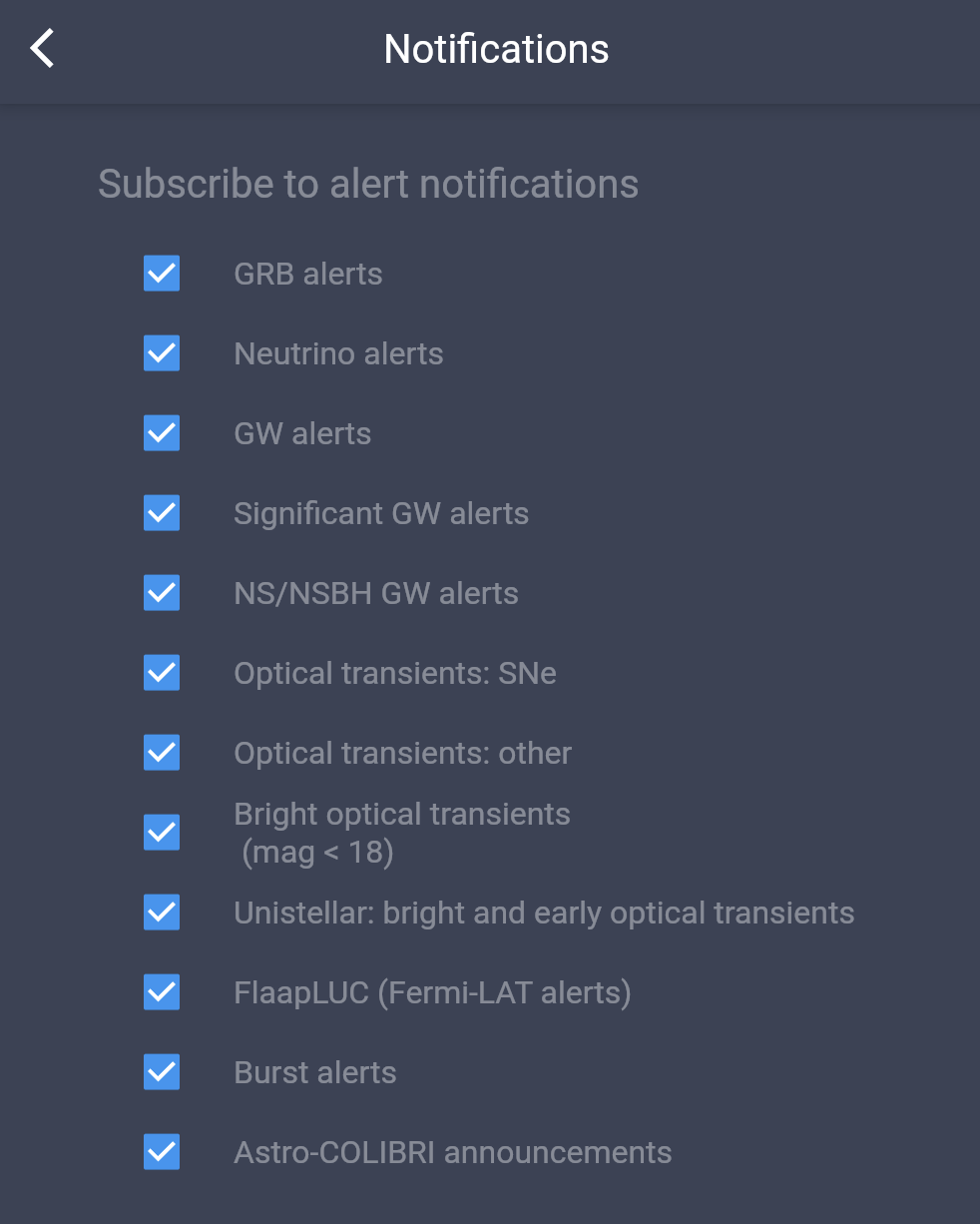}
\caption{Selection of realtime notification streams.}
\label{fig:notifications}
    \end{minipage} \hfill
    \begin{minipage}{0.4\textwidth}
        \centering
        \includegraphics[width=\textwidth]{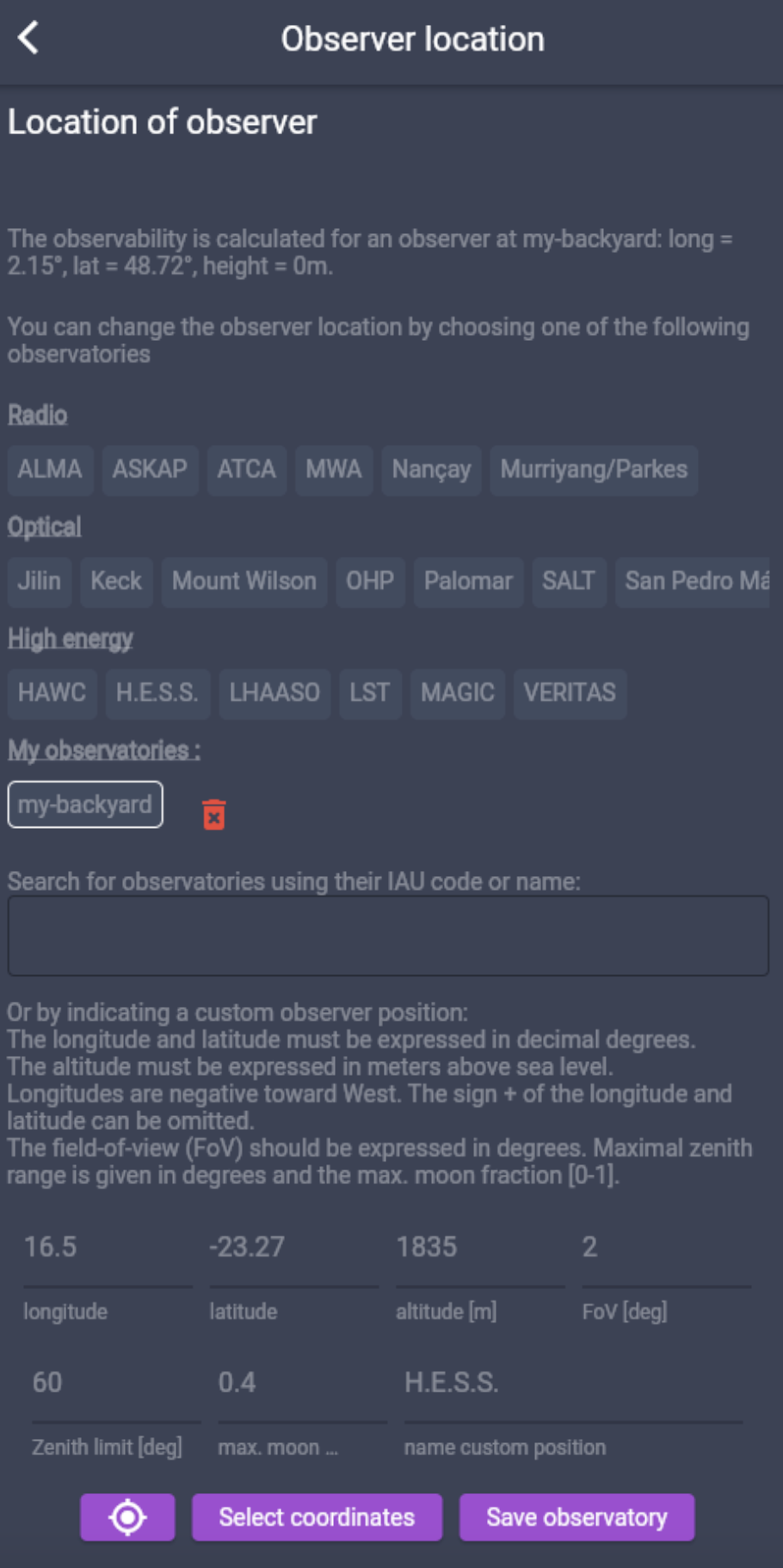} 
        \caption{Selection of follow-up observatories.}\label{fig:observatory}
    \end{minipage}
\end{figure}

One of the most prominent features of Astro-COLIBRI is the possiblity to receive alerts about new detections of transient phenomena in realtime on any smartphone. The free Android\footnote{\url{https://play.google.com/store/apps/details?id=science.astro.colibri}} and iOS\footnote{\url{https://apps.apple.com/us/app/astro-colibri/id1576668763}} apps can be downloaded from the relevant app stores by anyone interested in following time domain astrophysics news. At first launch of the app, several onboarding screen will be displayed to guide the user through the app. The first such screen is interactive and allows for the selection of the various available notification streams. These are served by the Google Firebase Cloud Messaging system and allow the recepetion of notifications in realtime. The available choices (v2.6.0 of the platform) are shown in Fig.~\ref{fig:notifications}. The same selection screen can also be found in the app menu, accessible via a button in the top right corner of the app.

Particularly interesting for amateur astronomers are the notification streams announcing the classification of bright optical transients. These are:
\begin{description}
    \item[Bright optical transients (mag$<$18)] Announcements of optical transients (SNe and other classified optical transients) with magnitudes $<$ 18 at the time of the detection of the new source. Thus, the source evolution since this moment is not taken into accocunt directly. The notification stream is a subset of two more extensive categories, which announce supernovae and all other optical transients independent of their magnitudes.
    \item[Unistellar: bright and early optical transients] These notifications announce the detection of candidate supernovae by the ZTF observatory and selected by the Unistellar Citizen Science program. Their magnitudes at the time of detection are usually required to be $<$ 16.3. These events are thus accessible by relatively small aperture telescopes. The selection procedure by the ALeRCE broker has large uncertainties due to the limited amount of information (e.g. typically only a single flux measurement exceeding the archival observations), the fraction of the events due to mis-classifications can thus be sizable.
\end{description}
    
\section{Observatory selection}
An important step towards follow-up observations of a particular event is the assessment of its observability at a given location that can be defined via the "Observer location" menu entry. This can be found in the app menu in the top right corner and on the web interface via the teardrop icon in the "personalize" area in the top row. Illustrated in Fig.~\ref{fig:observatory}, three complementary ways of selecting the observer location are available:

\begin{description}
    \item[Pre-defined observatories: ] Grouped by wavelength range, a set of professional observatories around the globe are available for rapid selection. 
    \item[Searches in the IAU database: ] Using a text field input with autocompletion, the user can search the database of the International Astronomical Union for additional observatories. 
    \item[Custom observatories: ] Most useful for citizen scientists and amateur astronomers, custom locations can be defined. The user can either rely on the GPS position of his device or input the coordinates and parameters of his/her setup manually. A user account (that can be created for free only with a valid email address) is necessary to save the created entries for later use.
\end{description}

\section{Gravitational waves}
Gravitational waves (GWs) and the associated multi-wavelength emissions allow unprecedented studies but are challenging due to the large localisation uncertainty regions provided by the gravitational wave interferometers. Follow-up observations aiming to localise the source of the GW emission thus require dedicated and optimized scheduling procedures. One such scheduling tool is {\it tilepy}~\citep{tilepy_icrc2023}, a packaged that has been developed within the H.E.S.S. collaboration~\citep{2021JCAP...03..045A}. The code is publicly available (cf. \href{https://github.com/astro-transients/tilepy}{tilepy@GitHub}). In addition, Astro-COLIBRI is providing an open, cloud-based and ready-to-use installation at \url{https://tilepy.com} from which the scheduling calculation can be requested easily via an API endpoint. 

\begin{figure}[t!]
    \centering
     \begin{minipage}{0.55\textwidth}
        \centering
        \includegraphics[width=\textwidth]{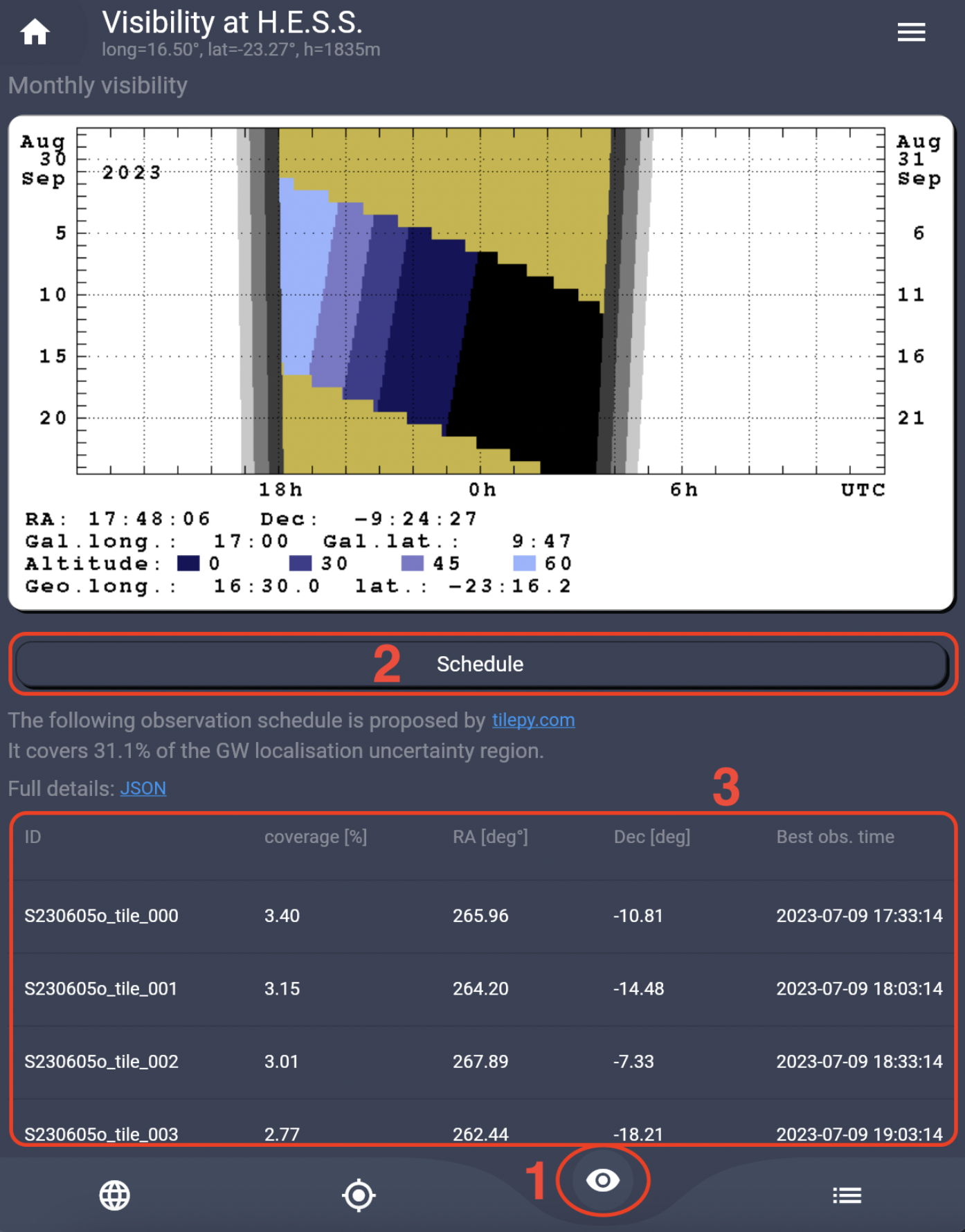} 
        \caption{Request of a GW follow-up schedule in the mobile Astro-COLIBRI app.}\label{fig:tilepy}
    \end{minipage} \hfill
    \begin{minipage}{0.4\textwidth}
        \centering
        \includegraphics[width=\textwidth]{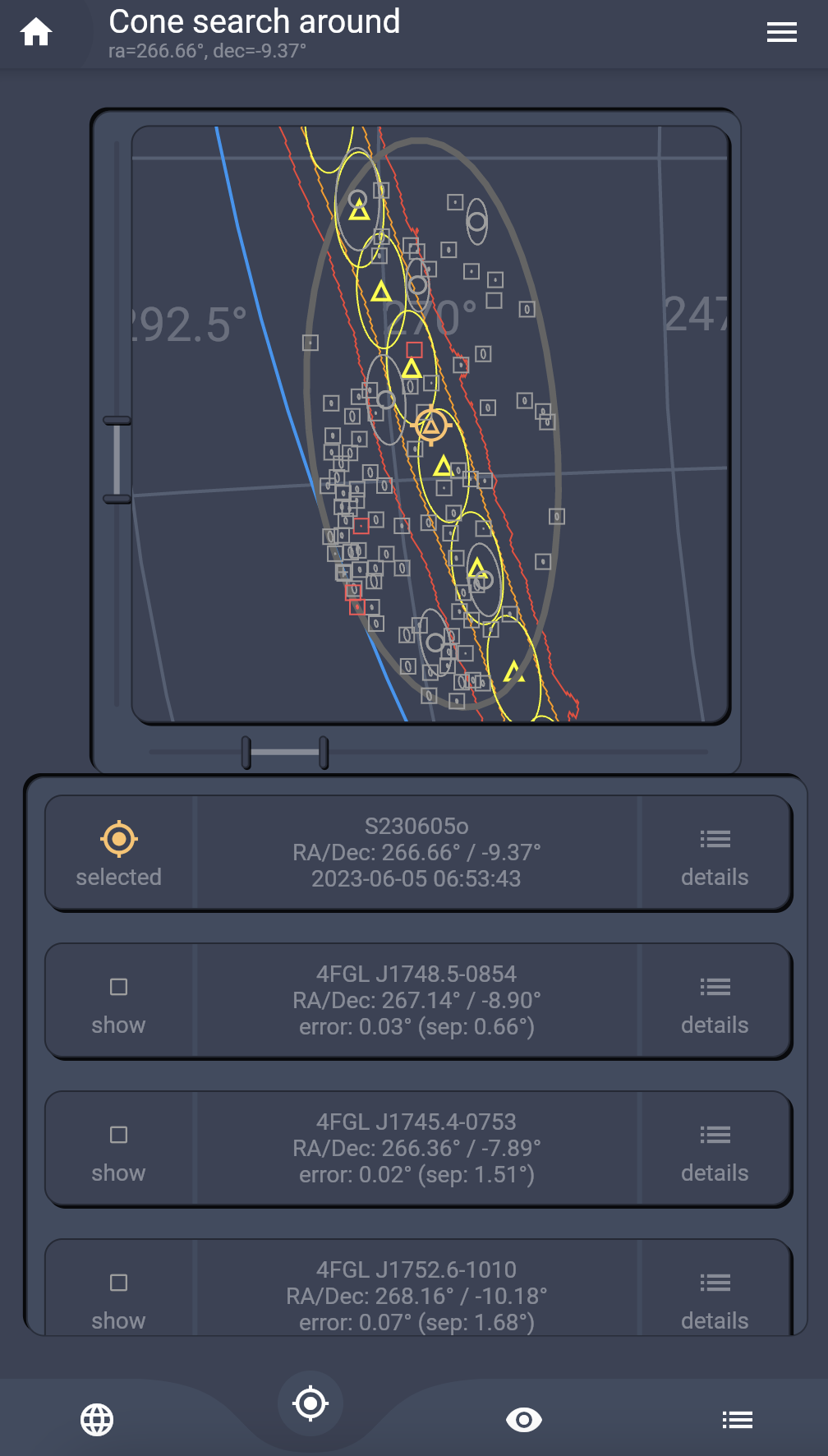} 
        \caption{Visualisation of the resulting observation plan.}\label{fig:tilepy_results}
    \end{minipage}
\end{figure}

Furthermore, the {\it tilepy} API has been integrated into the Astro-COLIBRI graphical interfaces. A typical use case can be described as: if matching the various user-defined filter criteria, Astro-COLIBRI users are notified upon the detection of a new GW event in real-time. They can verify all details and parameters of the event directly in the app. The localization of the event is displayed via two contours, representing 50 and 90\% containments. For each GW event and each user-selected follow-up observatory, an optimized follow-up schedule can then be obtained directly in the app or via the web interface (cf. Fig.~\ref{fig:tilepy}). The request can be submitted via a dedicated button (2) on the visbility screen (1). The derived observation positions are then automatically displayed in a table (3) and overlayed on the skymap, which allows one to check for possible related detections of transient events (e.g. a GRB, a high-energy neutrino, etc.). An example of this display is shown in Fig.~\ref{fig:tilepy_results}. Finally, the details of the observation plan can be downloaded in JSON format.

\section{Conclusions}
Astro-COLIBRI offers a comprehensive and efficient solution for real-time multi-messenger astrophysics, enhancing the discovery potential and collaboration within the astronomy community and with citizen scientists around the world. By integrating various data resources and providing intuitive interfaces, Astro-COLIBRI enables amateur and professional astronomers to better understand and explore the nature of transient astrophysical events.

The Astro-COLIBRI development team welcomes comments and feedback from the community to further improve the platform and can be contacted at \href{mailto:astro.colibri@gmail.com}{astro.colibri@gmail.com}.

\section{Acknowledgements}
The authors acknowledge the support of the French Agence Nationale de la Recherche (ANR) under reference ANR-22-CE31-0012. This work was also supported by the Programme National des Hautes Energies of CNRS/INSU with INP and IN2P3, co-funded by CEA and CNES and we acknowledge support by the European Union’s Horizon 2020 Programme under the AHEAD2020 project (grant agreement n. 871158).

\bibliographystyle{aa}  
\bibliography{sf2a-template} 

\end{document}